\documentclass[conference]{IEEEtran}
\IEEEoverridecommandlockouts
\usepackage{newtxtext,newtxmath}
\usepackage{cite}
\usepackage{url}
\usepackage{booktabs}
\usepackage{amsmath,amsfonts}
\usepackage{algorithm, algorithmic}
\usepackage{graphicx}
\usepackage{xcolor}
\usepackage{xurl}
\usepackage{balance}
\usepackage{csquotes, multirow, float, placeins, stfloats}
\def\BibTeX{{\rm B\kern-.05em{\sc i\kern-.025em b}\kern-.08em
    T\kern-.1667em\lower.7ex\hbox{E}\kern-.125emX}}

\begin{document}

\title{What You See Is Not Always What You Get: \\ Evaluating GPT's Comprehension of Source Code}

\author{
\IEEEauthorblockN{Jiawen Wen}
\IEEEauthorblockA{\textit{The University of Sydney}\\
Sydney, Australia\\
jwen8784@uni.sydney.edu.au}
\and
\IEEEauthorblockN{Bangshuo Zhu}
\IEEEauthorblockA{\textit{The University of Sydney}\\
Sydney, Australia\\
bzhu4004@uni.sydney.edu.au}
\and
\IEEEauthorblockN{Huaming Chen}
\IEEEauthorblockA{\textit{The University of Sydney}\\
Sydney, Australia\\
huaming.chen@sydney.edu.au}
}

\maketitle

\begin{abstract}

Recent studies have demonstrated outstanding capabilities of large language models (LLMs) in software engineering tasks, including code generation and comprehension. While LLMs have shown significant potential in assisting with coding, LLMs are vulnerable to adversarial attacks. In this paper, we investigate the vulnerability of LLMs to imperceptible attacks. This class of attacks manipulate source code at the character level, which renders the changes invisible to human reviewers yet effective in misleading LLMs' behaviour. We devise these attacks into four distinct categories and analyse their impacts on code analysis and comprehension tasks. These four types of imperceptible character attacks include coding reordering, invisible coding characters, code deletions, and code homoglyphs. To assess the robustness of state-of-the-art LLMs, we present a systematic evaluation across multiple models using both perturbed and clean code snippets. Two evaluation metrics, model confidence using log probabilities of response and response correctness, are introduced. The results reveal that LLMs are susceptible to imperceptible coding perturbations, with varying degrees of degradation highlighted across different LLMs. Furthermore, we observe a consistent negative correlation between perturbation magnitude and model performance. These results highlight the urgent need for robust LLMs capable of manoeuvring behaviours under imperceptible adversarial conditions. 
 
\end{abstract}


\section{Introduction}
The success of Large Language Models (LLMs) in natural language processing (NLP) has driven their widespread adoption across many different sectors, one of which is software development. The presented ability to understand and generate code snippets through LLMs makes them highly accessible, resulting in the integration into developers' workflows, including code generation \cite{Du24}, program repair \cite{Zhang23}, and vulnerability detection \cite{Zhou24}.

However, the human-like responses of LLMs do not constitute as true reasoning capabilities, and their integration into software engineering workflows introduces new security risks and vulnerabilities. Considering their increasing adoption and continued growth in their sector, it is important to ensure the robustness of LLMs against tempted adversarial attacks and study their response mechanisms. Understanding these weaknesses in LLMs is fundamental to the future LLM-based software development.

Recently, there are several studies demonstrating the feasibility of text-based adversarial attacks against LLMs, targeting performance degradation of NLP tasks \cite{Boucher22,Formento2021,Li20}. While these studies have successfully featured the LLMs vulnerabilities, it remains unknown concerning the feasibility of these attacks in real-world scenarios. Dyrmishi et al. argues that for an adversarial attack to be practical against a text-based LLM, the generated text must be both valid and natural to human reviewers~\cite{Dyrmishi23}. For attacks that fail to preserve high semantic consistency or produce text that appears artificially generated, the corresponding real-world effectiveness will be limited. Through a survey of 378 human participants, their study concludes that existing text-based adversarial attacks lack imperceptibility, limiting their effectiveness in real-world scenarios where human reviewers are involved.

To address this research gap, this paper presents a novel class of \textit{imperceptible adversarial attacks} on source code, specifically targeting LLMs in code analysis and comprehension tasks. Our goal is to investigate how the LLMs' behaviour changes when presented with subtly manipulated code that appears identical to human reviewers. We define \enquote{true imperceptibility} as adversarial code snippet that is visually and semantically indistinguishable from its original form. While prior work such as~\cite{Liu21} explores imperceptible attacks in NLP, those manipulations often leave detectable artifacts while may be semantically and visually \textit{similar}. Our work demonstrates the feasibility and effectiveness of \textit{truly imperceptible}, character-level attacks in realistic software engineering scenarios, bridging the gap between theoretical vulnerabilities analysis and practical threats.


Thus, in this work, our attack method introduces a set of perturbation techniques to source code with the goal of achieving \textit{true imperceptibility}. Specifically, we leverage the special Unicode specification characters that subtly affect visual rendering while preserving readability for human reviewers. Inspired by Boucher et al. \cite{Boucher22}, we design four distinct imperceptible attack strategies targeting code analysis and comprehension tasks, namely code reordering, invisible coding characters, code deletions, and code homoglyphs. We include different LLMs solutions, specifically with the state-of-the-art GPT models, for the attack performance evaluation. The efficacy of these attacks are assessed based on two key performance metrics in a large-scale code analysis task, which are \textit{model confidence} and \textit{response correctness}. Experiment results demonstrates a strong negative correlation between perturbation and performance outcomes, with some notable differences among these LLMs, which we will further elaborate in Section \ref{sec:results}.

In summary, this paper makes the following contributions:

\begin{enumerate}
    \item We investigate the vulnerability of LLMs to imperceptible adversarial attacks in the context of code analysis and comprehension tasks with LLMs.
    \item We propose a black-box imperceptible coding character attack framework that introduces four distinct attack strategies against the state-of-the-art LLMs solutions. The strategies includes code reordering, invisible coding characters, code deletions, and code homoglyphs, which effectively manipulate LLMs' behaviour while remaining visually imperceptible for code analysis and comprehension task.
    \item Extensive experiments are conducted to demonstrate the effectiveness of the attacks, covering a wide range of dataset and LLMs. We highlight the correlation between the attack budget and the performance outcomes.
    \item We have released the replication package including the codebase and results\footnote{Replication package: https://figshare.com/s/71c3544e89d4a9fe8a61} to facilitate the future development and research of LLMs for coding tasks. 
\end{enumerate}

The remaining sections of the paper are as follows. Section~\ref{sec:motivation} presents the motivation and background of this study. We provide the details of the overall framework design and experiment setup in Section~\ref{sec:studydesign}, and report the experimental results in Section~\ref{sec:results}. Section~\ref{sec:discussion} discusses our findings with novel insights, and Section~\ref{sec:conclusion} summarises the paper.

\section{Motivation and Background}
\label{sec:motivation}

\subsection{LLM for Coding Tasks}

Large Language Models (LLMs) have achieved strong performance across a broad range of natural language processing tasks, including sentiment analysis \cite{Ye09}, text classification \cite{Dogra22}, and scientific writing. In high-stakes domains such as healthcare, LLMs have been evaluated for clinical reasoning, summarisation, and patient guidance \cite{Cascella23}. Despite these advances, LLMs are prone to hallucination, producing fluent yet factually incorrect outputs, and are often deployed without formal guarantees of robustness or security.

The adoption of LLMs in software engineering is emerging, with tools like GitHub Copilot\footnote{https://github.com/features/copilot} and Amazon Q Developer\footnote{https://aws.amazon.com/q/developer/} integrating LLMs for code generation and review. Codex, a GPT-based model trained on large-scale public code repositories, exemplifies this trend by generating Python code from docstrings and natural language input \cite{Chen21}. Studies have also explored LLMs' capacity to explain and analyse code snippets, which forms the basis for tasks like static code analysis and review automation \cite{Fang24}.

\subsection{Imperceptible Perturbations and Technical Gap}

Although adversarial prompt attacks have been widely studied in natural language processing, including synonym substitution and syntactic rewrites \cite{Ye09, Dogra22}, these techniques are typically visible and less applicable to source code tasks where minor character-level changes can affect semantics. In software engineering, especially in code review or comprehension, perturbations that are invisible to humans but still affect machine interpretation pose a unique and underexplored threat.

Boucher et al. \cite{Boucher22} introduced a new class of attacks, termed imperceptible NLP attacks, which exploit the visual similarity or invisibility of certain Unicode characters. Their taxonomy included four categories: homoglyph substitutions, reordering of characters, invisible characters, and deletions. These perturbations degrade model performance on various NLP tasks without altering visible output. The study also demonstrated real-world implications, including bypassing content filters and degrading search engines, by misaligning human and machine understanding of the same string.

Table \ref{tab:imperceptible-attack} illustrates typical examples adapted from the imperceptible attack class. For instance, right-to-left override (\enquote{U+202E}) and repeated backspace characters (\enquote{U+8}) are invisible in most editors but significantly alter the semantics of the input.

\begin{table}[!t]
\caption{Examples of imperceptible attacks adapted from \cite{Boucher22}}
\label{tab:imperceptible-attack}
\centering
\renewcommand{\arraystretch}{1.2}
\begin{tabular}{|c|c|}
\hline
\textbf{Input encoding} & \textbf{UI rendering} \\
\hline
photo\_high\_re\textcolor{red}{U+202E}gnp.js & photo\_high\_res.png \\
\hline
I do not\textcolor{red}{U+8}\textcolor{red}{U+8}\textcolor{red}{U+8}\textcolor{red}{U+8}\textcolor{red}{U+8}\textcolor{red}{U+8} authorise this & I authorise this \\
\hline
\end{tabular}
\end{table}

While Fang et al.\cite{Fang24} examined LLM behaviour under code obfuscation techniques such as variable renaming or restructuring, their work did not address character-level visual deception. Our work builds on the foundations laid by Boucher et al., extending the scope from NLP tasks to source code understanding and vulnerability analysis in LLMs.

\subsection{Real-World Attack Scenarios and Overtrust}

\textbf{Non-Expert Users.} An increasing number of users, including hobbyists, students, scientists, copy and reuse code from LLM-generated outputs, online repositories, or technical forums. Lacking formal software engineering expertise, these users increasingly rely on the same LLMs (e.g., ChatGPT, Copilot Chat) to validate these code snippets \cite{arxiv2401, acmNonExpert2024, danielZeroCode2023, arxiv2211insecure}. This creates a critical risk: perturbed or adversarially crafted code may be incorrectly interpreted as valid, with users unaware of the underlying syntactic or security implications of what they are copying.

\textbf{Overtrust and Automation Bias.} Many studies report that users tend to overtrust LLM outputs, even when flawed or speculative, \cite{cacmCopilot2025, Fang24, overtrust2025, arxivTrustAI2024}. For instance, in healthcare, parents have shown a preference for ChatGPT responses over advice from certified physicians \cite{kuParents2024}. In software engineering, such overtrust can manifest as the uncritical adoption of flawed code analyses or faulty model outputs. 

\textbf{Organisational Integration and Supply Chain Risk.} We have seen enterprises are increasingly integrating LLMs into CI/CD pipelines and code auditing workflows, such as Amazon Q Developer and Copilot. If imperceptible attacks evade detection by LLM-powered validation layers, these malicious codes can be introduced into the software supply chain with internal security checks undetected \cite{storek2025xoxo, infosecAICode2024, paloaltoAICyber2023, ibmAIOps}. Without adversarial hardening~\cite{dyrmishi2023empirical}, the implicit trust placed in LLMs within development operations can become a significant security liability.

\section{Framework and Experiment Design}
\label{sec:studydesign}

\begin{figure*}[t]
    \centering
    \includegraphics[width=\textwidth]{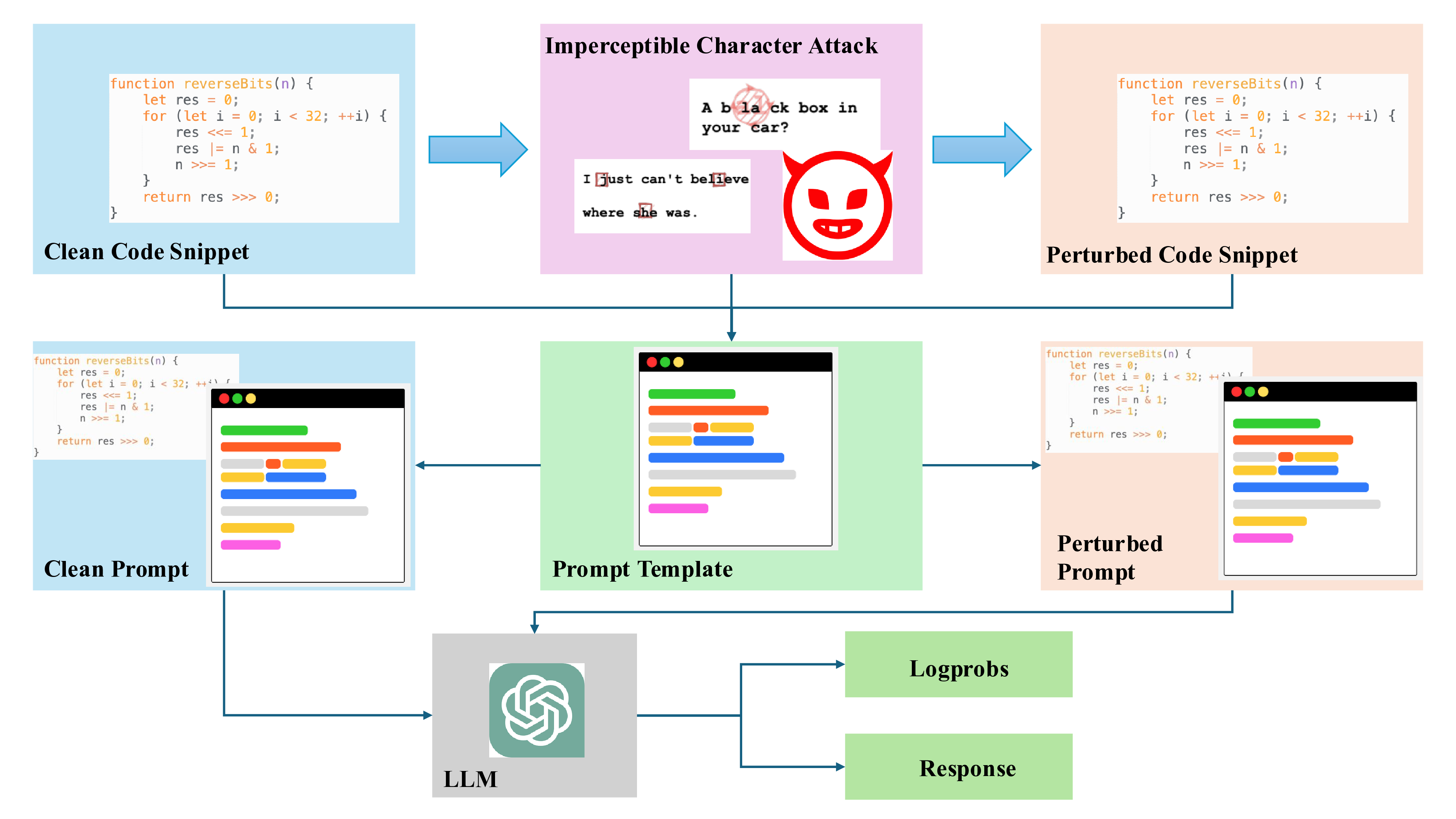}
    \caption{Overall imperceptible coding character attack framework}
    \label{attackProcess}
\end{figure*}

In this work, we aim to evaluate the effectiveness and identify the limitations of imperceptible coding character attacks on LLMs in the context of code comprehension tasks. Following we will outline the details of proposed research questions, overall black-box imperceptible coding character attack framework, and the experiment design.

\subsection{Research questions}

To guide the investigation, this study seeks to answer the following four questions:

\begin{itemize}
    \item \textbf{RQ1:} To what extent do imperceptible perturbations within code correlate with LLMs' performance outcomes in software testing scenarios?
    \item \textbf{RQ2:} Does a correlation exist between imperceptible perturbations and LLMs' \enquote{confidence}, and what implications does this have for software testing?
    \item \textbf{RQ3:} How significantly do different imperceptible perturbation methods impact the responses of LLMs?
    \item \textbf{RQ4:} How do advancements in recent LLMs development improve their robustness against imperceptible adversarial perturbations in software testing scenarios?
\end{itemize}

\subsection{Methodology}

Figure~\ref{attackProcess} presents the overall attack framework used in this study. We adopt a unified prompting strategy composed of three key elements:

\begin{enumerate}
    \item Code snippet
    \item Description of the code snippet
    \item A binary question: \enquote{Does the provided code match the provided description? Answer with either Yes or No.}
\end{enumerate}

The OpenAI Chat Completions API \cite{chatcomp} is used to submit each prompt to the target LLM. For each input, the code snippet is perturbed according to a specific strategy while the accompanying description remains unchanged. This setup isolates the effect of code-only perturbations on the model’s understanding and judgment.

The model’s response is constrained to a single-token binary answer: either “Yes” or “No”. This design choice ensures that the evaluation remains fully quantifiable and reproducible, avoiding the ambiguity inherent in open-ended responses. By enforcing a binary format, we enable direct, consistent comparison of correctness across all test cases and facilitate precise use of the \enquote{logprobs} property \cite{logprobs} to measure model confidence. Additionally, this format mitigates the need for human interpretation, which would be required to judge multi-token explanations and could introduce subjective bias and inconsistency—particularly problematic in large-scale or automated evaluations. Finally, the Yes/No structure also reflects practical real-world scenarios, where users often pose concise security-related questions to LLMs, such as “Is this code safe?”

The entire response generation process is fully automated, requiring no human involvement for response validation or grading. To ensure scalability and objectivity, we designed the experiment to produce correctness and confidence scores programmatically, without relying on model self-evaluation or human annotation. The next section describes the specific perturbation strategies used to challenge the model's comprehension capabilities.

\subsection{Attack taxonomy}

Herein, we discuss our solution for the imperceptible coding character attack design, which devises four categories of perturbation design, each featuring a special Unicode character implementation:

\begin{itemize}
    \item \textbf{Code reordering:} Character U+202E is used so the Unicode specification can support characters from languages that read from right-to-left \cite{202e}. Inserting it at the head and tail of the reversed string means that to the human eye, the perturbed section is rendered in the correct order, but the input will be the reversed string.
    \item \textbf{Invisible coding characters:} Character U+200C is a character that does not render to a visible glyph. When placed between two characters that would otherwise be joined by a ligature, it keeps them disconnected. It is also effectively a space character.
    \item \textbf{Code deletions:} Character U+0008 is the backspace character. It can be injected into a string to remove characters from visual rendering.
    \item \textbf{Code homoglyphs:} Homoglyphs are unique character that render to the same glyph. The Unicode Consortium publishes a document with the Unicode Security Mechanisms technical report \cite{intentional} which details a mapping of characters that are intended as homoglyphs within the Unicode specification. These characters are intended to be mapped to the same glyph in font implementations and are thus \enquote{true} homoglyphs. The technical report also includes a document which details a set of characters that are commonly visually confused, although these glyphs are not intended to be mapped to the same glyph in font implementations \cite{unintentional}. Since this second set is dependent on the choice of font implementation, and the extent to which they are visually similar is subjective, they were not considered for this study.
\end{itemize}

\subsection{Attack generation}

To increase the control of the experiment, the decision was made to create perturbed samples programmatically, rather than generatively. This means that the data perturbation process is open-source and consistent, rather than generating them using a process that both involves closed-source software and produces inconsistent output. The perturbation process could not be exactly replicated across each category of perturbation, but each process shared in common the following 5 traits:

\begin{enumerate}
    \item Each perturbation is rendered identically to the clean code.
    \item The perturbations are inserted starting from beginning to end.
    \item The levels of perturbation budget are set at 20\%, 40\%, 60\%, 80\%, and 100\%.
    \item The perturbations are created with special Unicode characters, but not exclusively so.
    \item The floor is taken for any mathematical operation involving non-integer numbers.
\end{enumerate}

Algorithm~\ref{alg:perturb} formalises this perturbation process across four perturbations, providing detailed design to generate adversarial examples for downstream evaluation. Further details can be found in the replication package.

\begin{algorithm}[H]
\caption{Imperceptible Code Perturbation Process}
\label{alg:perturb}
\begin{algorithmic}[1]
\REQUIRE Clean code $C$ (length $L$), perturbation type $T \in \{\texttt{REORDER}, \texttt{INVISIBLE}, \texttt{DELETE}, \texttt{HOMOGLYPH}\}$, perturbation budget $b \in [0,1]$
\ENSURE Perturbed code $C'$

\STATE $n \gets \lfloor b \cdot L \rfloor$
\STATE $C' \gets$ empty string

\IF{$T = \texttt{REORDER}$}
    \STATE prefix $\gets \text{reverse}(C[0{:}n])$
    \STATE suffix $\gets C[n:]$
    \STATE $C' \gets \text{unicodeReverse}(\text{prefix}) + \text{suffix}$
    \RETURN $C'$
\ENDIF

\STATE $count \gets 0$
\FOR{$i = 0$ to $L - 1$}
    \STATE $c \gets C[i]$
    
    \IF{$T = \texttt{INVISIBLE}$ \AND $count < n$}
        \STATE $C' \gets C' + c + \texttt{U+200C}$
    
    \ELSIF{$T = \texttt{DELETE}$ \AND $count < n$}
        \STATE $C' \gets C' + \texttt{"a"} + \texttt{U+0008} + c$
    
    \ELSIF{$T = \texttt{HOMOGLYPH}$ \AND $count < n$}
        \STATE $C' \gets C' + \text{homoglyph}(c)$
    
    \ELSE
        \STATE $C' \gets C' + c$
    \ENDIF

    \STATE $count \gets count + 1$
\ENDFOR

\RETURN $C'$
\end{algorithmic}
\end{algorithm}

\subsection{Prompting}

We craft the prompts with the zero-shot prompting technique. We ensure that the templates of each prompt remain consistent across all categories of perturbations and all models. Following the implementation of different attack strategies, we leverage the \textit{Chat Completions API} provided by the OpenAI to access the tested LLMs, which can be found in the replication package. The prompt template consists of the following three messages:

\begin{enumerate}
    \item Question
    \item Code
    \item Description
\end{enumerate}

\noindent
The question remains constant across all prompts, and reads:

\begin{displayquote}
    \textit{Does the provided code match the provided description? Answer with either Yes or No.}
\end{displayquote}

We note a similar question was explored in the study of \cite{Fang24}, which is open-ended in nature. However, our methodology adopts a more constrained approach by rephrasing the question to elicit a one-word response. Moreover, both the code and accompanying descriptions in our study are sourced from the LeetCode dataset~\cite{huggingface}. As such, we treat the description as the \enquote{ground truth}, given the curated process of the dataset. The perturbations are introduced to the code that preserve its overall semantics from a human perspective. This experiment design ensures that the correct answer to the question is consistently \enquote{Yes}, allowing for a controlled and reliable evaluation of LLMs' comprehension. 

\subsection{Logprobs}

To present a thorough data analysis, we used the \enquote{logprobs} parameter of the Chat Completions API as the main vector. With logprobs enabled, the Chat Completions API returns the log probabilities of each output token in the response. Log probabilities of output tokens indicate the likelihood of each token occurring in the sequence \cite{logprobs}. A higher log probability suggests a higher likelihood of a token in that context, which allows the confidence of the LLM in its response to be gauged. 

To ensure meaningful and fair interpretation of these values, we have constrained the model's output to a single token, which is either \enquote{Yes} or \enquote{No}. It guarantees that the log probability of the response could provide meaningful data. This experimental design provides two aspects of the response:

\begin{enumerate}
    \item The response will be exactly one token long
    \item The response will be either \enquote{Yes} or \enquote{No}
\end{enumerate}

Since there is no variation in token length, our experiments control for the confounding effects introduced by multi-token outputs. In fact, this constraint is motivated by observations from a preliminary study of the experiment, in which unconstrained responses resulted in inconsistent and noisy logprobs results due to contextual dependencies across multiple tokens. It proves impossible to distinguish meaningful tokens from superfluous ones, prompting the adoption of this one-token constraint. 

Given that the LLM must respond with either \enquote{Yes} or \enquote{No}, and the correct answer is always \enquote{Yes} which is guaranteed by the curated dataset, the response enable two key measurements:

\begin{enumerate}
    \item The correctness of the response (Corr.\%).
    \item The confidence of the LLM in this response (Conf.\%).
\end{enumerate}

\subsection{Confidence Score Calculation}

To evaluate model confidence, we utilise the log probabilities returned by the OpenAI Chat Completions API with the \texttt{logprobs=true} setting. For each response, we convert the log-scale values into linear probabilities using the exponential function, average them across all tokens, and scale the result to a percentage range. This yields a normalised confidence score that reflects the model's average token-level certainty in generating its full response.

Let \( \log P_i \) be the log probability of the \( i \)-th token in a response, and let \( N \) be the total number of tokens. The confidence for a given response is computed as:
\[
\text{Confidence} = \left( \frac{1}{N} \sum_{i=1}^{N} e^{\log P_i} \right) \times 100
\]

To measure the degradation caused by perturbations, we center each perturbed response’s confidence around the average confidence of clean (unperturbed) responses. For a perturbed response, we define the sample-level adjusted score as:

\[
\text{Score}_{\text{perturb}} = 
\begin{cases}
\text{Conf}_{\text{perturb}} - \text{AvgCleanConf}, & \text{if correct} \\
-100, & \text{if incorrect}
\end{cases}
\]

This scheme penalises incorrect predictions sharply, ensuring they strongly influence the average. For correct responses, the score reflects relative changes in confidence without distorting scale.

Finally, to report confidence under each perturbation level in the results tables, we re-center the aggregate score by adding back the average clean confidence:

\[
\text{DisplayedConf}_{\text{perturb}} = \text{AvgCleanConf} + \frac{1}{N} \sum_{i=1}^{N} \text{Score}_{\text{perturb}}^{(i)}
\]

This formulation ensures that the reported confidence remains on the same scale as clean outputs, while still reflecting both prediction accuracy and confidence degradation under perturbation. Confidently incorrect answers are heavily penalised, while correct but uncertain answers contribute proportionally.

\subsection{Selected LLMs System}

In this study, we select three different GPT models for evaluation. In order of recency:

\begin{itemize}
    \item \textbf{gpt-3.5-turbo-0613:} A snapshot of the GPT-3.5 turbo model released in June 2023, used to represent an earlier iteration of the GPT-3.5 generation.
    \item \textbf{gpt-3.5-turbo-0125:} A more recent variant of GPT-3.5 turbo released in January 2024, included to reflect incremental improvements within the same generation.
    \item \textbf{gpt-4o-2024-05-13:} A GPT-4 series model representing a newer generation with enhanced performance characteristics \cite{gpt4o}. This allows for comparative analysis between 3.5-series models and a more advanced architecture in handling code perturbations.
\end{itemize}

\subsection{Dataset}

To identify a large-scale dataset suitable for this study, inspiration is taken from a similar study \cite{Jalil23} which prompted ChatGPT with software-related questions. In that study, ChatGPT is tasked with answering questions from a \enquote{popular software testing textbook}. While this approach offers insightful findings, it is limited in scale for textbook questions. Differently, we adopt a similar solution to a broader and more scalable dataset, particularly with a greater number of code snippets.

Herein, we leverage the LeetCode dataset from huggingface.co \cite{huggingface}, which is an AI community website featuring user-submitted models and datasets. It provides a rich set of code-description pairs for large-scale evaluation. In this work, we utilize a dataset of 2,644 LeetCode coding questions. Given the curated nature of LeetCode, the code-description pairs are assumed to be reliable and trustworthy, without requiring a \enquote{base truth} for code snippets. The questions in the dataset have three categories of difficulty: easy, medium, and hard. Each question has answers available in four different coding languages: Java, C++, Python, and JavaScript. For this experiment, only JavaScript code snippets are used, to control for the confounding variable of different languages. 

\section{Results}
\label{sec:results}

\begin{figure}[htp]
    \centering
    \includegraphics[width=1\linewidth]{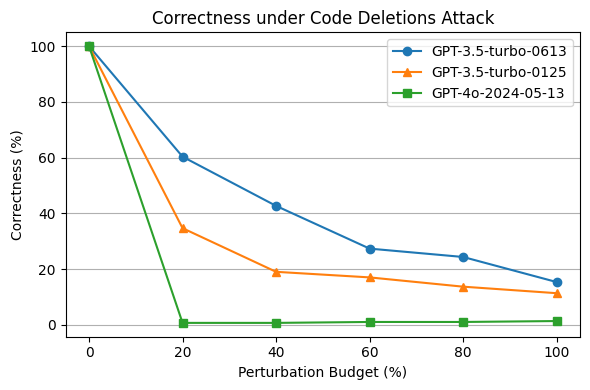}
    \caption{Correctness vs.\ perturbation budget for the \textit{Code Deletions} attack across three models.}
    \label{fig:result_overview}
\end{figure}

\subsection{Overview}



We investigated the robustness of three GPT-based language models, two GPT-3.5 turbo variants (GPT-3.5-turbo-0613 and GPT-3.5-turbo-0125) and one GPT-4o version (GPT-4o-2024-05-13), against four code perturbation methods: reordering, invisible characters, deletions, and homoglyphs. Our findings show that all models consistently achieve 100\% correctness on unaltered (clean) code, demonstrating their baseline ability to comprehend straightforward snippets. However, the GPT-3.5 models generally display a uniform, step-wise decline in confidence and correctness as the perturbation budget increases, particularly for deletions and reordering, while invisible characters and homoglyphs prove slightly less disruptive. In contrast, GPT-4o exhibits a drastic performance drop as soon as any level of perturbation is introduced, yielding near-zero correctness in most cases and indicating that the newer model may have additional filtering or checks that invalidate the perturbed code. These trends highlight how different perturbation types undermine model confidence at varying rates and underscore the stark contrast in failure modes between the GPT-3.5 and GPT-4o generations.

\subsection{Results on GPT-3.5-turbo-0613}

From the Table \ref{tab:3.5-turbo-0613}, we can immediately find many similarities across the 4 categories. As expected, the confidence score for the clean code snippet is high with none below 95\%, and the correctness is 100\% across the board. The highest confidence score is held by invisible characters at 96.11\%. Each category has their largest drop in confidence between clean and 20\% perturbation budget, with differences getting smaller as the perturbation budget rises. The largest difference between budget levels is held by the deletions category between clean and 20\%, at -52.76\%. Reorderings and deletions have significantly lower floors for both confidence and correctness scores compared to invisible characters and homoglyphs, both approaching 0\% for confidence, and both sub-20\% for correctness. Deletions holds the lowest score across all perturbation budget levels for both confidence at 1.74\% as well as correctness at 15.3\%. The highest scores at 100\% perturbation budget are held by the homoglyph category, with confidence at 54.23\% and correctness at 67.67\%. The homoglyph category is also the only perturbation category to not show decreases in scores at every level, with an increase in both confidence and correctness scores recorded between 60\%-100\%.

\begin{table*}
\centering
\caption{Performance of the model \textit{GPT-3.5-turbo-0613} under four imperceptible coding character attack methods.}
\begin{tabular}{l|cc|cc|cc|cc}
\toprule
\multirow{2}{*}{\textbf{Perturbation budget}} 
  & \multicolumn{2}{c|}{\textbf{Reordering}} 
  & \multicolumn{2}{c|}{\textbf{Invisible characters}} 
  & \multicolumn{2}{c|}{\textbf{Deletions}} 
  & \multicolumn{2}{c}{\textbf{Homoglyphs}} \\
\cmidrule(lr){2-3}\cmidrule(lr){4-5}\cmidrule(lr){6-7}\cmidrule(lr){8-9}
 & \textbf{Conf.\%} & \textbf{Corr.\%}
 & \textbf{Conf.\%} & \textbf{Corr.\%}
 & \textbf{Conf.\%} & \textbf{Corr.\%}
 & \textbf{Conf.\%} & \textbf{Corr.\%} \\
\midrule
\textbf{Clean}   & 95.47 & 100 & 96.11 & 100 & 95.45 & 100 & 95.74 & 100 \\
\textbf{20\%}    & 54.81 & 72.67 & 51.80 & 65.67 & 41.84 & 60.33 & 77.85 & 86.67 \\
\textbf{40\%}    & 38.42 & 54.67 & 42.85 & 58.00 & 20.65 & 42.67 & 58.47 & 69.33 \\
\textbf{60\%}    & 26.91 & 47.33 & 35.15 & 50.67 & 10.11 & 27.33 & 49.93 & 63.00 \\
\textbf{80\%}    & 11.28 & 27.67 & 33.36 & 51.00 &  7.42 & 24.33 & 51.33 & 68.00 \\
\textbf{100\%}   &  3.62 & 16.33 & 29.64 & 47.67 &  1.74 & 15.33 & 54.23 & 67.67 \\
\bottomrule
\end{tabular}
\label{tab:3.5-turbo-0613}
\end{table*}

\subsection{Results on GPT-3.5-turbo-0125}

The results for this model displays a similar profile to the model previous, with all categories sharing characteristics in common. The clean code is 100\% correct and has the highest confidence score across the board, with all categories at 93±1\%. The highest confidence score is held by invisible characters at 93.84\%. Each category has their largest drop in confidence between clean and 20\% perturbation budget, with differences getting smaller as the perturbation budget rises. The largest difference between budget levels is again held by the deletions category between clean and 20\%, at -74.67\%. Another aspect held in common between the two GPT-3.5 models are reorderings and deletions having significantly lower floors for both confidence and correctness scores compared to invisible characters and homoglyphs, with both sub-10\% for confidence, and both sub-25\% for correctness. Deletions again holds the lowest overall confidence score at 3.93\%, but at 60\% budget rather than 100\% budget for this more recent model. The highest scores at 100\% perturbation budget are still held by the homoglyph category, with confidence at 49.80\% and correctness at 64.33\%. A difference to the older model is that here both the homoglyph category and the invisible characters category do not show decreases in scores at every level, with an increase in both confidence and correctness scores recorded between 80\%-100\% for both.

\begin{table*}
\centering
\caption{Performance of the model \textit{GPT-3.5-turbo-0125} under four imperceptible coding character attack methods.}
\begin{tabular}{l|cc|cc|cc|cc}
\toprule
\multirow{2}{*}{\textbf{Perturbation budget}} 
  & \multicolumn{2}{c|}{\textbf{Reordering}} 
  & \multicolumn{2}{c|}{\textbf{Invisible characters}} 
  & \multicolumn{2}{c|}{\textbf{Deletions}} 
  & \multicolumn{2}{c}{\textbf{Homoglyphs}} \\
\cmidrule(lr){2-3}\cmidrule(lr){4-5}\cmidrule(lr){6-7}\cmidrule(lr){8-9}
 & \textbf{Conf.\%} & \textbf{Corr.\%}
 & \textbf{Conf.\%} & \textbf{Corr.\%}
 & \textbf{Conf.\%} & \textbf{Corr.\%}
 & \textbf{Conf.\%} & \textbf{Corr.\%} \\
\midrule
\textbf{Clean}   & 93.51 & 100   & 93.84 & 100   & 93.69 & 100   & 93.46 & 100 \\
\textbf{20\%}    & 28.55 & 46.33 & 48.61 & 66.33 & 19.02 & 34.67 & 75.16 & 84.00 \\
\textbf{40\%}    & 18.80 & 36.33 & 44.16 & 62.67 &  8.05 & 19.00 & 58.91 & 72.67 \\
\textbf{60\%}    & 15.94 & 34.33 & 35.71 & 54.67 &  3.93 & 17.00 & 49.71 & 64.33 \\
\textbf{80\%}    & 11.83 & 26.33 & 31.03 & 48.33 &  4.80 & 13.67 & 43.94 & 60.67 \\
\textbf{100\%}   &  8.44 & 22.00 & 37.38 & 56.67 &  4.60 & 11.33 & 49.80 & 64.33 \\
\bottomrule
\end{tabular}
\label{tab:3.5-turbo-0125}
\end{table*}

\subsection{Results on GPT-4o}

Table \ref{tab:gpt4o} displays a clear and significant difference to the two GPT-3.5 turbo models. However, there is one characteristic held in common across the results of all three models. Across all 4 categories, the clean code is still 100\% correct and still holds the highest confidence score, with all of them holding a score of 81±1\%. However, from there, things change drastically. The GPT-4o results are the only results to show negative confidence scores, which is due to the disproportionate number of incorrect answers, with the lowest score of -19.08\% held by reorders at 20\% perturbation budget. As with confidence scores, correctness is also extremely low across the board, with the next highest score following 100\% being 2.33\%, also held by reorders at 100\% perturbation budget. The lowest correctness score is tied between reorders at 20\% perturbation budget and invisible characters at 100\% perturbation budget, and is also the lowest possible correctness, 0\%.

\begin{table*}
\centering
\caption{Performance of the model \textit{GPT-4o-2024-05-13} under four imperceptible coding character attack methods.}
\begin{tabular}{l|cc|cc|cc|cc}
\toprule
\multirow{2}{*}{\textbf{Perturbation budget}} 
  & \multicolumn{2}{c|}{\textbf{Reordering}} 
  & \multicolumn{2}{c|}{\textbf{Invisible characters}} 
  & \multicolumn{2}{c|}{\textbf{Deletions}} 
  & \multicolumn{2}{c}{\textbf{Homoglyphs}} \\
\cmidrule(lr){2-3}\cmidrule(lr){4-5}\cmidrule(lr){6-7}\cmidrule(lr){8-9}
 & \textbf{Conf.\%} & \textbf{Corr.\%}
 & \textbf{Conf.\%} & \textbf{Corr.\%}
 & \textbf{Conf.\%} & \textbf{Corr.\%}
 & \textbf{Conf.\%} & \textbf{Corr.\%} \\
\midrule
\textbf{Clean}   & 80.92 & 100  & 81.46 & 100  & 80.94 & 100  & 81.28 & 100 \\
\textbf{20\%}    & -18.35 & 0.67 & -17.05 & 1.33 & -18.29 & 0.67 & -17.54 & 1.00 \\
\textbf{40\%}    & -19.08 & 0    & -16.34 & 2.00 & -18.31 & 0.67 & -17.58 & 1.00 \\
\textbf{60\%}    & -18.59 & 0.67 & -17.43 & 1.00 & -18.27 & 1.00 & -18.00 & 0.67 \\
\textbf{80\%}    & -18.82 & 0.33 & -17.80 & 0.67 & -17.92 & 1.00 & -18.35 & 0.33 \\
\textbf{100\%}   & -17.24 & 2.33 & -18.54 & 0    & -17.52 & 1.33 & -18.36 & 0.33 \\
\bottomrule
\end{tabular}
\label{tab:gpt4o}
\end{table*}

\subsection{Answer to research questions}

\noindent\textbf{RQ1: To what extent do imperceptible perturbations within code correlate with LLMs' performance outcomes in software testing scenarios?}


To answer this question, we take the term \enquote{performance outcomes} to refer to the correctness of a model’s responses. The introduction of perturbations can be found to greatly affect the correctness of the model responses, which can be gleaned simply from looking at the tables containing the summary of results. Statistical analysis confirms this conclusion using Pearson’s correlation coefficient, with a moderate-to-strong negative correlation coefficient found between perturbation budget and response correctness for all perturbation categories, as shown in Table \ref{correctness}. Simply put, this means that as perturbation goes up, correctness goes down proportionally. The two GPT-3.5 generation models have very similar results, with both appearing to exhibit significantly higher negative correlation than the GPT-4o model. If only considering the correlation coefficient, it seems as if the GPT-4o model also shows a very high degree of normalization between the four categories of perturbation compared to the two 3.5 models. However, knowing that the results show a distribution discrepancy between the GPT-3.5 generation models and the GPT-4o model, we can investigate further.

\begin{table}[h!]
\centering
\caption{Correlation coefficient of correctness scores against perturbation budget}
\renewcommand{\arraystretch}{1.3}
\begin{tabular}{c l l l l}
 \hline
 & \textbf{Reord.} & \textbf{Invis.} & \textbf{Del.} & \textbf{Homog.} \\ 
 \hline
 \textbf{GPT-3.5-turbo-0613} & -0.98 & -0.87 & -0.94 & -0.90 \\
 \textbf{GPT-3.5-turbo-0125} & -0.85 & -0.91 & -0.83 & -0.66 \\
 \textbf{GPT-4.0 with clean considered} & -0.64 & -0.67 & -0.65 & -0.97 \\
 \textbf{GPT-4.0 with clean discarded} & 0.64 & -0.85 & 0.94 & -0.95 \\
 \hline
\end{tabular}
\label{correctness}
\end{table}

If we consider the full range of perturbation budgets, from 0\% to 100\%, the GPT-4o model results exhibit a correlation coefficient of approximately -0.6 across all categories, indicating a moderately strong negative correlation. However, when we exclude the clean code and focus only on cases where the perturbation budget is greater than 0\%, we observe a strong positive correlation in the reordering and deletion categories, while the remaining categories continue to show a strong negative correlation.

\begin{table}[h!]
\centering
\caption{Number of correct responses (out of 1500) across different GPT models and perturbation types}
\label{tab:model_prop}
\renewcommand{\arraystretch}{1.3}
\begin{tabular}{lccccc}
\hline
\textbf{Model} & \textbf{Clean} & \textbf{Reord.} & \textbf{Invis.} & \textbf{Del.} & \textbf{Homog.} \\
\hline
\textbf{GPT-4o} & 1500 & 12 & 15 & 14 & 10 \\
\textbf{GPT-3.5-turbo-0613} & 1500 & 656 & 819 & 510 & 1064 \\
\textbf{GPT-3.5-turbo-0125} & 1500 & 496 & 866 & 287 & 1038 \\
\hline
\end{tabular}
\end{table}

Table \ref{tab:model_prop} shows the proportion of total correct responses to incorrect responses returned by the GPT-4o model across all perturbation budgets. Since each category of perturbation creates five different levels of perturbation, and this study works with a 300-length dataset, each category of perturbation will make 1500 total prompts. Additionally, since the clean code is prompted with every perturbed code snippet, rather than only once per code-description pair, the clean prompt is made 1500 times too. The results reveal that for the GPT-4o model responses, any correct response is in fact an outlier as the number of correct responses does not represent a significant proportion of the total number of responses, and it is the reason for the contradictory correlation coefficient exhibited by the statistical analysis.

For reference, Table \ref{tab:model_prop} also shows the proportion of total correct responses to incorrect responses returned by the two GPT-3.5 models. It is immediately obvious that for these models, there is a statistically significant number of correct responses for each category. This information also explains why the correlation coefficient for GPT-4o is so similar across all four categories of perturbation when the control is considered, whilst the GPT-3.5 models exhibit a much higher degree of variance. However, despite these limitations in GPT-4o results, statistical analysis can still conclusively demonstrate that the performance outcomes of the model are affected by perturbation, even if we cannot find a linear correlation between the correctness scores against the perturbation budget levels. \\

\noindent\textbf{RQ2: Does a correlation exist between imperceptible perturbations and LLMs' \enquote{confidence}, and what implications does this have for software testing?}

Similarly to the correctness scores, the results from Table \ref{confidence} exhibits shows a clear negative correlation between confidence scores and perturbation, which is again corroborated by statistical analysis using Pearson’s correlation coefficient. Likewise, there is a clear discrepancy between the results of the GPT-3.5 generation models and the GPT-4o model when clean responses are discarded, with the underlying reason being the same as for RQ1.

\begin{table}[H]
\caption{Correlation coefficient of confidence scores against perturbation budget}
\renewcommand{\arraystretch}{1.3}
\centering
\begin{tabular} { c l l l l }
 \hline
 & \textbf{Reord.} & \textbf{Invis.} & \textbf{Del.} & \textbf{Homog.} \\ 
 \hline
 \textbf{GPT-3.5-turbo-0613} & -0.72 & -0.55 & -0.72 & -0.50 \\
 \textbf{GPT-3.5-turbo-0125} & -0.58 & -0.49 & -0.61 & -0.53 \\
 \textbf{GPT-4.0 with clean considered} & -0.64 & -0.65 & -0.64 & -0.65 \\
 \textbf{GPT-4.0 with clean discarded} & 0.01 & -0.01 & 0.01 & -0.01 \\
 \hline
\end{tabular}
\label{confidence}
\end{table}

What we can infer from these results is largely the same as what was inferred in the discussion of RQ1. GPT-3.5 models exhibit clear negative correlation between confidence scores and perturbation budget, meaning that as perturbation goes up, confidence goes down proportionally. Correspondingly, for GPT-4o we can only say that the confidence of the model is affected by perturbation, but we cannot demonstrate a linear correlation between the confidence scores against perturbation budget levels.\\

\noindent\textbf{RQ3: How significantly do different imperceptible perturbation methods impact the responses of LLMs?}

\begin{figure}[h]
    \centering
    \includegraphics[width=0.4\textwidth]{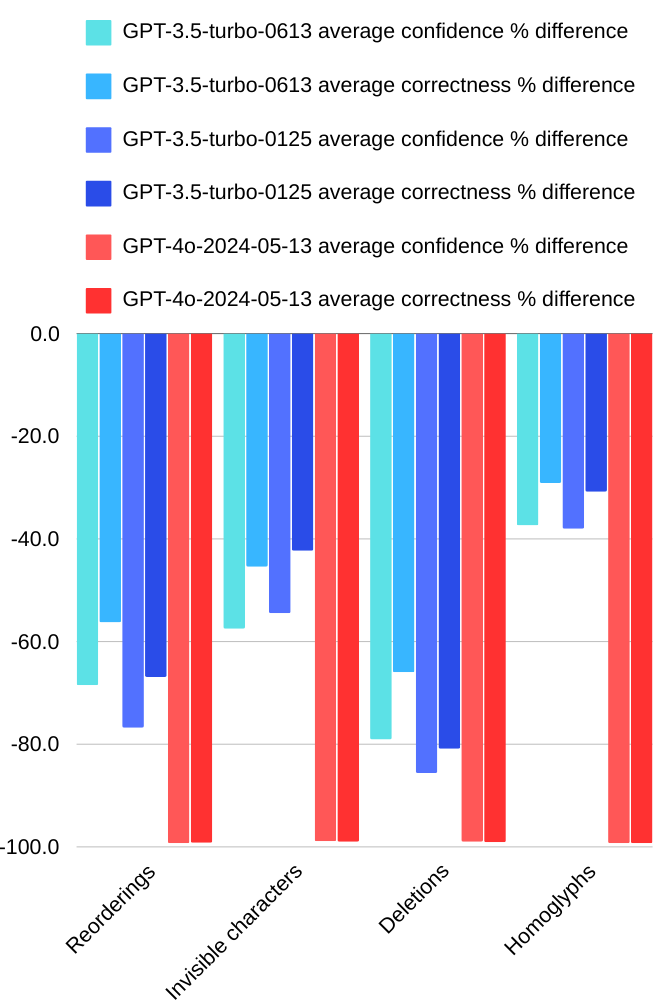}
    \caption{Average difference in confidence scores for all models}
    \label{avgDiff}
\end{figure}

To answer this question, we must look at the difference between the clean control response scores, and the perturbed response scores. Figure \ref{avgDiff} records the average difference in response score across all perturbation budgets for each category of perturbation. We can clearly see that deletion perturbations had the greatest overall effect, and homoglyphs had the lowest overall effect. The homoglyph category being the lowest is easily explained by the fact that there are simply a smaller proportion of characters affected for each perturbation when compared to the other three categories, as homoglyphs calculate their perturbation budget against the set of all possible homoglyphs. This subset is always smaller in size than the set of all characters of the original string, which is what the three other categories calculate their perturbation budget against.

As with the previous research questions, the GPT-4o model contradictory results. All categories show close to -100\% average difference for both overall confidence and overall correctness scores. This means that the overwhelming majority of responses were incorrect, previously established by Table \ref{tab:model_prop}. We can also conclude that GPT-4o does not differentiate between categories of perturbation, nor perturbation budgets. And the results are the same whether 20\% perturbed or 100\% perturbed.\\

\noindent\textbf{RQ4: How do advancements in recent LLMs development improve their robustness against imperceptible adversarial perturbations in software testing scenarios?}

A trend throughout the results suggests significant internal differences between the GPT-3.5 generation models and the GPT-4o generation model. We can deduce that some deterministic guardrails have been put in place for the GPT-4o generation model which the GPT-3.5 generation models do not have. Whilst the GPT-3.5 models lose confidence and correctness proportionally to the amount of perturbation present, GPT-4o returns an incorrect response across the board when presented with any category of perturbation, at any budget level.

However, even though the response is classified as \enquote{incorrect} by the criteria of the study, it does not denote that the incorrect response is an undesirable outcome. A naïve conclusion drawn from the results of this study would be that the older generation of models are simply better at comprehending perturbed code than the newest generation, but the limitations of the study design mean that we cannot determine for what reason the GPT-4o model returned a \enquote{No} answer. The more useful conclusion we should draw from the results is that the GPT-4o model can detect (with very high accuracy) the presence of special characters, recognises their potential to confound the performance outcomes, and then deterministically returns a response that guarantees no false positives will be presented to the user.

\section{Discussion}
\label{sec:discussion}

\subsection{Novel insights}

The purpose of this study is to contribute to the depth of research in large language model perturbation, rather than the breadth. Critical insight is gained into the inner mechanisms of three iterations of the most popular large language model in use, which has been widely adopted in certain security-sensitive sectors whilst being closed-source.

The process of undertaking this study has also provided insight into how the experiment methodologies of subsequent studies should be designed. Knowing now that the \enquote{flagship} OpenAI product \cite{gpt4o} exhibits behaviour that indicates the existence of deterministic guardrails put in place between the 3.5 and 4 generation models, we hope that ensuing research can investigate the finer details of such guardrails, and their effects on a system that is designed to be interacted with using natural language. Guardrails alone can negate all false positives, but a truly sophisticated system should recognise the presence of confounding characters, parse the code according to the spirit of the prompt, and return a correct answer, nonetheless; there should be no correlation between perturbation and correctness at all. However, in the absence of such a sophisticated system, it is encouraging to see that the evolution from 3.5-generation GPT to 4-generation GPT brings with it an unequivocally more consistent model.

\subsection{Contributions and implications}

In this study, we devise four different categories of perturbation on three different LLM models. For the two older generation models, a moderately strong negative linear correlation was found between the level of perturbation and the performance outcomes of the response. The newer generation model appears to exhibit behaviours that indicate the existence of deterministic guardrails.

The perturbation category of deletions was found to have the highest overall effect on performance outcomes, followed by reordering, invisible characters, and finally by homoglyphs. The results of homoglyphs are not entirely conclusive as the study methodology calculates their perturbation budget differently to the other three categories; further investigation can be made.

Findings of the study further establish the baseline effectiveness of LLMs in code comprehension tasks, with all 18 000 clean code prompts returning responses with the correct answer and a high internal confidence score.

The methodology of this study draws on practices of preceding studies. However, efforts have been made to reduce the amount of qualitative analysis in data gathering, so that results are not hampered by the subjective confounding variables inherent to human analysis. Particular emphasis is placed on not gathering data on model confidence through model self-reporting, instead, using quantitative tools to provide evidence in this regard. This study made use of the \enquote{logprobs} property of the Chat Completions API so that conclusions regarding model confidence can be drawn from quantitative data rather than qualitative model self-reporting – it is the author’s hope that subsequent studies will implement similar quantitative analysis procedures regarding confidence.

\subsection{Assumptions and limitations}

\textbf{We assume a \enquote{No} response means the LLM does not \enquote{understand} the code at all}. This study was designed with the assumption that a \enquote{No} response is a strong indication that the large language model does not understand the code at all. The \enquote{ideal behaviour} of the model in the event of an imperceptible prompt injection attack was defined as it being able to parse it as a human does, i.e. ignore or \enquote{reverse-engineer} the perturbed code and parse the cleaned version. However, the results of the study, especially the discrepancy between the GPT-3.5 models and the GPT-4o model, show that this projected \enquote{ideal behaviour} is of dubious import, and that it may rather be better to create a system which can guarantee no false negatives, rather than only being able to answer a subset of adversarial examples.

\textbf{Log probabilities are a good indicator of confidence}. A secondary objective of this research has been to harden data gathering procedures against confounding variables which are exhibited in extant research. LLM self-evaluation has been used in preceding studies to record LLM confidence scores, but an explicit requirement in the methodology design of this study is to avoid self-reporting to side-step potential confounding variables inherent in relying upon a closed-source system to evaluate itself. However, it may be the case that log probabilities are also not a perfect representation of model confidence.

What the log probabilities represent is the probability a token is chosen to appear in a specific context \cite{logprobs}. For example, imagine a model is prompted with the straightforward question: \enquote{Is the sky blue?} and it returns a single-token response \enquote{Yes}. This token has a log probability which converts to a linear percentage value of 90\%. What that log probability tells us is that this hypothetical model had a 90\% chance to respond with \enquote{Yes}, and a 10\% chance to respond with anything else. Now, let us imagine that we prompt the model with a more subjective question such as: \enquote{Is the Maserati Quattroporte a beautiful car?} and it again returns a single-token response \enquote{Yes}. The log probability of that response will be lower, as it is more likely that the model responds \enquote{No}, even if it ultimately still returned the same response as before. We take this difference between the likelihood of an answer to represent the fall in confidence of the model. Whilst this practice is supported by official documentation \cite{logprobs}, independent research disagrees. One such example is a paper by Panickssery et al. \cite{Panickssery24} which indicates that LLMs recognise and favour their own generations even when making evaluations on log probabilities of response tokens. This is why the overall assumption is included here, as a large part of the validity of the gathered data rests on the assumption that log probabilities are an accurate representation of model confidence.

\textbf{The confidence scores of wrong answers are recorded as \enquote{-100}}. In the design of this study, in the event the response is different from what we know the base truth to be, the confidence score of that response is recorded as \enquote{-100}. While it becomes easier to draw conclusions on overall correctness, and confidence scores in correct responses, it normalises the data in a way which makes it difficult to draw conclusions on the causes of incorrect responses. This design was made with the assumptions outlined in the previous section in mind, and new insights into the veracity of those assumptions may affect the validity of the design of this study.

\textbf{Heavy limitations placed on responses}. The design of this study limits responses to two possible one-token-long responses: either \enquote{Yes} or \enquote{No}. This limitation was enacted so that there would be fewer confounding variables at play when performing log probability confidence evaluation. Preliminary runs of this study design that did not limit token length of response found very high variance in linear probability of responses, as the log probability of each token in the response had to be averaged. This meant that no useful data was gathered with regards to model confidence by using log probabilities, and it would have necessitated human analysis of qualitative results. Since quantitative analysis was greatly preferred to cut down on confounding variables, the decision was made to limit the response length and response options.

\subsection{Potential Threats and Defense Strategies}

While imperceptible character-based perturbations may sometimes prevent code from compiling, our goal is not to evaluate the correctness of the code snippets. Furthermore, it does not eliminate their security relevance for the investigation of LLMs' capability, especially in workflows where LLMs are used to review or validate code before it is ever executed. We highlight two key threat scenarios and propose defense strategies that address them at various stages of the LLM-assisted code analysis pipeline.

\textbf{Threat 1: Post-copy vulnerability propagation.} In real-world workflows, code is often copied between platforms (e.g., from a website to an IDE, or from a chat to a version control system). During this process, certain imperceptible characters may be stripped or normalised, resulting in a clean-looking version of the code that compiles and runs, but one that was previously reviewed and falsely approved by an LLM. This creates a false sense of safety, as neither the LLM nor the human reviewer recognised the original obfuscation.

\textbf{Threat 2: User confusion and resource waste.} When perturbed code does survive intact, developers may face unclear compilation errors or behavioural inconsistencies. These issues can lead to time-consuming debugging, misdiagnosed problems, or abandonment of useful code. In high-pressure environments, such disruptions erode trust in AI code review tools and increase reliance on manual inspection.

To mitigate these risks, we recommend the following defense strategies:

\textbf{1. Pre-processing and normalization:} Implement input sanitisation steps such as Unicode normalisation, removal of zero-width characters, and visual similarity checks (e.g., homoglyph detection). These measures reduce the likelihood of LLMs encountering malformed or deceptive input.

\textbf{2. Multi-pass verification:} Avoid relying solely on a single LLM judgment. Instead, aggregate results from multiple models or rerun the same model under varied conditions. Only proceed with critical decisions when consensus or confidence thresholds are met.

\textbf{3. Explainability enforcement:} Require LLMs to justify their reasoning when labelling code as safe or unsafe. Structured explanations can help detect hallucinated judgments, expose oversights, and provide transparency for both humans and downstream tools.

These strategies are not intended to guarantee complete robustness. Rather, they offer a layered defense framework tailored to the unique risks posed by imperceptible code perturbations in LLM-centered pipelines. Addressing these subtleties is essential for building trustworthy, secure AI-assisted development environments.

\subsection{Future work}

The primary limitation of this study lies in finding a way to quantitatively assess model confidence. It is suggested to use novel and effective practices or tools in the future that can better guarantee the validity of confidence scores to produce data from which concrete conclusions can be drawn.

Future research should also delve into how we can create LLM-based systems that can generate a correct response even if the prompt exhibits certain level of perturbations. This is to bridge the gap between what the user experiences and expect, and what the model truly \enquote{understands}. Guardrails have been shown to negate false positives entirely, which is a marked improvement upon previous generations of LLMs. However, a truly sophisticated natural language processor should be able to follow the spirit of a prompt, rather than the letter of it.

\section{Conclusion}
\label{sec:conclusion}

The findings in this paper identify the vulnerabilities in recent LLMs for imperceptible coding character attack. With the black-box imperceptible coding character attack framework, we have introduced four distinct attack strategies that effectively manipulate the LLMs' behaviour in coding related tasks. Our primary goal is to assess whether LLMs can maintain alignment with the original code intent with minimal visual perturbations, which perturbation may not result in functionally executable code snippets. We have conducted large-scale experiments to demonstrate the susceptibility of LLMs against code snippet perturbations. Furthermore, we have observed that perturbation budget had a strong negative correlation to performance outcomes for code analysis and comprehension task. Additionally, concerning the latest LLMs solution, there is a strong negative correlation between the presence of perturbation and performance outcomes, but no valid correlational relationship between perturbation budget and performance outcomes. We claim that the current devised solutions of imperceptible coding character attacks have presented significant impacts for the coding tasks. We anticipate future research will focus on bridging such issue to enhance the trust of LLMs, in particular for code analysis and comprehension tasks.


\textbf{Data Availability:} To promote open science policy, we have fully released our code, experimental process, experimental results, which is now available at the following link: \url{https://figshare.com/s/71c3544e89d4a9fe8a61}.

\balance
\bibliographystyle{IEEEtran}
\bibliography{paper}

\end{document}